\documentclass[9pt,conference]{IEEEtran}

\usepackage[utf8]{inputenc}
\usepackage{graphicx} 
\usepackage{float} 
\usepackage{caption} 
\usepackage{subcaption} 
\usepackage{todonotes} 
\usepackage{wrapfig} 
\usepackage{rotating} 
\usepackage{verbatim} 
\usepackage{mathtools} 
\usepackage{latexsym} 
\usepackage{listings} 
\usepackage[linesnumbered]{algorithm2e} 
\usepackage{amsfonts} 
\usepackage{balance} 
\usepackage{eso-pic} 
\usepackage[pdftex,
            pdfauthor={Jan Novacek, Ali Ahari, Alessandro Cornaglia, Frederik Haxel, Alexander Viehl, Oliver Bringmann, Wolfgang Rosenstiel},
            pdftitle={Ontology-supported Design Parameter Management for Change Impact Analysis},
            pdfsubject={This paper presents an ontology-supported approach to the management of design parameters in engineering.  This approach aims specifically at enabling  Change Impact Analysis through Requirements Traceability and acquainted expert knowledge of design parameters. The approach is suitable for both software and hardware designs. The activities and features are mainly obtained by (1) the application of an ontology-based universal system modeling procedure proposal for model integration, (2) the utilization of a knowledge base for capturing expert knowledge and (3) a semantic Mission Profile Aware Design platform. OWL is used to represent information and the underlying data model can improve knowledge transfer among heterogeneous systems which are common in complex engineering projects. At the same time,  effort to perform reasoning on such models can be reduced. A demonstration and hands-on description of two illustrative use cases complements the paper.},
            pdfkeywords={Systems Engineering, Knowledge Based Engineering, Requirements Management, Mission Profiles},
            pdfproducer={LaTeX with hyperref},
            pdfcreator={pdflatex},
            bookmarks=false,
            colorlinks=true,
            citecolor=black,
		    filecolor=black,
		    linkcolor=black,
		    urlcolor=black]{hyperref}
\usepackage[square,sort,comma,numbers]{natbib} 
\usepackage{amsmath} 
\newcommand{\IEEEAcceptedManuscriptNotice}{%
  \AddToShipoutPictureFG*{%
    \AtPageLowerLeft{%
      \put(48,12){%
        \parbox[b]{516pt}{%
          \hrule
          \vspace{2pt}
          \fontsize{5.2}{5.8}\selectfont
          \raggedright
          \textcopyright~2018 IEEE. Personal use of this material is permitted.
          Permission from IEEE must be obtained for all other uses, in any current
          or future media, including reprinting/republishing this material for
          advertising or promotional purposes, creating new collective works, for
          resale or redistribution to servers or lists, or reuse of any copyrighted
          component of this work in other works.\par
          \vspace{0.7pt}
          \textit{This is the accepted manuscript of: J. Novacek, A. Ahari,
          A. Cornaglia, F. Haxel, A. Viehl, O. Bringmann, and W. Rosenstiel,
          ``Ontology-Supported Design Parameter Management for Change Impact
          Analysis,'' in 2018 44th Euromicro Conference on Software Engineering and
          Advanced Applications (SEAA), Prague, Czech Republic, pp. 9--16, 2018.
          The version of record is available at
          \href{https://doi.org/10.1109/SEAA.2018.00011}{https://doi.org/10.1109/SEAA.2018.00011}.}%
        }%
      }%
    }%
  }%
}

\begin{document}
\IEEEAcceptedManuscriptNotice

\title{Ontology-supported Design Parameter Management\\for Change Impact Analysis}
%


\author{\IEEEauthorblockN{Jan Novacek\IEEEauthorrefmark{1}\IEEEauthorrefmark{2}, Ali Ahari\IEEEauthorrefmark{1}, Alessandro Cornaglia\IEEEauthorrefmark{1}, Frederik Haxel\IEEEauthorrefmark{1}\\ Alexander Viehl\IEEEauthorrefmark{1}, Oliver Bringmann\IEEEauthorrefmark{1}\IEEEauthorrefmark{2}, Wolfgang Rosenstiel\IEEEauthorrefmark{1}\IEEEauthorrefmark{2}}
\IEEEauthorblockA{\IEEEauthorrefmark{1}FZI Forschungszentrum Informatik\\
Haid-und-Neu-Str. 10-14, 76131 Karlsruhe}
\IEEEauthorblockA{\IEEEauthorrefmark{2}Eberhard Karls Universität Tübingen\\
Sand 14, 72076 Tübingen}}

\maketitle              

\begin{abstract}
This paper presents an ontology-supported approach to the management of design parameters in engineering.  This approach aims specifically at enabling  Change Impact Analysis through Requirements Traceability and acquainted expert knowledge of design parameters. The approach is suitable for both software and hardware designs. The activities and features are mainly obtained by (1) the application of an ontology-based universal system modeling procedure proposal for model integration, (2) the utilization of a knowledge base for capturing expert knowledge and (3) a semantic Mission Profile Aware Design platform. OWL is used to represent information and the underlying data model can improve knowledge transfer among heterogeneous systems which are common in complex engineering projects. At the same time,  effort to perform reasoning on such models can be reduced. A demonstration and hands-on description of two illustrative use cases complements the paper.
\end{abstract}

\section{Introduction}
\label{sec:introduction}
Requirements imposed on the development of systems frequently change and the consequences of such changes cannot be foreseen. Moreover, frequent changes are the reason why monitoring, traceability and Change Management mechanisms are needed. Requirements Traceability is especially important for the development of safety-critical systems and is required by standards such as the ISO 26262. The Change Impact Analysis (CIA) methodology proposed in this paper uses Trace Links and, in addition, acquainted expert knowledge about the relation between design parameters and requirements. Approaches that consider multiple scopes (as code, architecture and miscellaneous artifacts) are rare. As a consequence, Lehnert concluded that \emph{"[...] more attention should be paid on linking requirements, architectures and code to enable comprehensive impact analysis."} \cite{lehnert2011review}.

\begin{figure}
  \centering
  \includegraphics[width=\linewidth]{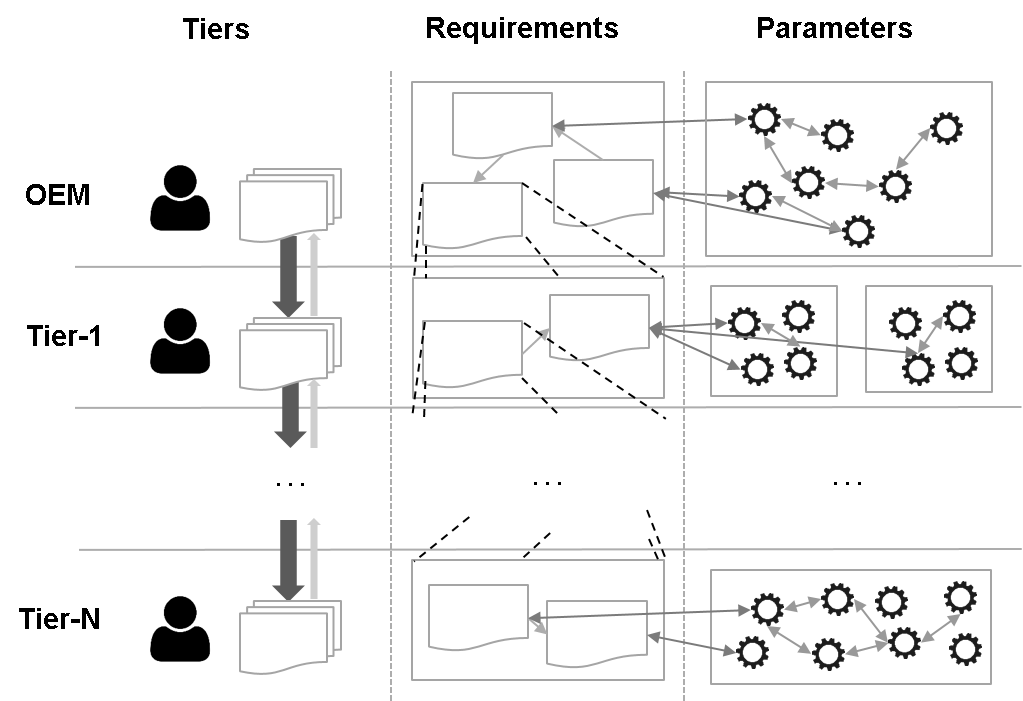}
  \caption[bla]{Supply chain with interdependencies between requirements and design parameters}
  \label{fig:requirements-passing-on}
\end{figure}

Typically, requirements are passed along the supply chain while being refined and broken down to separate tier needs, as shown in Fig. \ref{fig:requirements-passing-on}. Requirements can be linked to one or more \emph{design parameters} which in this context can basically be any characteristic of the artifact or system in question such as the range of a radar module, a system configuration setting or the runtime of a specific function of a software component. The design parameters themselves have interdependencies that are called \emph{operative relationships}. The range of a radar module is for example related to the gain of its amplifier.
%

Basically, this paper aims at providing support for decision making during development and uses expert knowledge about operative relationships. Employing Trace Links between requirements and design artifacts or representations thereof combined with acquainted expert knowledge about design parameters makes CIA easier and contributes to making the right decision with adequate effort. In addition, formalized knowledge of design parameters and ontological representations of systems are essential for reasoning on system models, e.g. validation, transformation or exchange. In addition, the approach considers Mission Profiles (MPs) in the design process thus enabling Mission Profile Aware Design (MPAD).

Operative relationships are mostly unknown. Therefore, consequences from requirement changes might not be taken into consideration and could cause problems due to undesired \emph{side effects} and/or \emph{ripple effects}. In addition, this is why CIA is complex and requires significant effort. As the everyday use of model-based approaches in the automotive and avionic industries is still limited, heterogeneous models and tools are hardly integrated which is identified as a key challenge in engineering settings by Broy et al. in \cite{broy2010seamless}.
Borg et al. observe in their comprehensive case study that manual work dominates CIA in industry \cite{borg2017ciaRecomm}.
This might be a reason for underestimated or unnoticed system impacts and incomplete or delayed analyses \cite{rovegaard2008empirical}.

To deal with these issues, the main purpose of the presented approach is the realization of a system that is able to provide an answer to the question "Which design parameters are affected if a requirement is changed?". Providing an answer to this question involves the integration of heterogeneous data sources.
Benefits are primarily:
\begin{itemize}
\item Avoidance of errors and improvement of overall product quality
\item Reduced costs due to the decreased complexity of a CIA
\item Improved understanding of the design and the relation between design parameters and requirements
\end{itemize}
The proposed \emph{system model} used in the approach is ontology-based and follows the principle of Linked Data that allows interlinking and discovery of new data sources in a global data graph \cite{bizer2009linked}. Relying on an established standard for ontology language (OWL \cite{w3dc2012owl}), the system also favors sharing and processing of system models used in development. Creating effective Knowledge Management Systems is a key success factor in engineering process improvement \cite{chourabi2008ontologySE}.

The main contribution of this paper is the proposed methodology for integrating heterogeneous data sources via the establishment of an ontology-based system model and the use of ontological representations. In addition to data integration applying this methodology enables specification of links between design parameters and requirements as well as traceability information.

This paper is divided into five sections. The Introduction is followed by a section providing relevant background information. The third section describes the solution approach and the fourth section presents two use cases that constitute a hands-on description. The final section consists of a conclusion drawn from the experiences made while carrying out this research.

\section{Background and Related Work}
\label{sec:related_work}
This section describes relevant background knowledge. An introduction to MPs and MPAD follows a comparison of the approach with existing work in Knowledge-based and Model-based Systems Engineering. This section also provides some insight to state-of-the-art of ontologies in Requirements Engineering. In addition, there is an outline of the CIA that is used in the presented approach comparing it to related work.

\subsection{Mission Profiles}
\label{sec:mission_profiles}
MPs are used to capture and specify environmental conditions to which components are exposed throughout their life cycle \cite{byrne2008handbook}. They contain various relevant stress factors and are therefore composed of different sources such as load usage and environmental profiles, requirements and scenarios for stress tests. MPs play an important role at the design and development of automotive systems, especially with regard to the propagation of requirements along the development process chain \cite{nirmaier2014mission,abelein2012complexity}. 
There is no common standard for MPs yet. A format draft to specify MPs has been developed in the RESCAR2.0 project\footnote{https://www.edacentrum.de/rescar/} and is further described in \cite{nirmaier2014mission}. The autoSWIFT project\footnote{https://www.edacentrum.de/autoswift/} aims at improving and standardizing the format.

\subsection{Mission Profile Aware Design}
The general concept of MPAD was introduced by Jerke and Kahng in \cite{jerke2014mission}. According to their paper, MP consideration is still mainly a manual task. Thus, providing tool support is essential for making adoption easier and enabling automation. The presented approach realizes this using the Mission Profile Framework and the semantic MPAD platform as described in Section \ref{sec:semantic_mpad_platform}. The Reliability Knowledge Framework by NXP contains a MP library and was created to connect users, reliability knowledge, data, tools and methods \cite{rongen2014reliabilityFramework}.
Today, apart from the mentioned frameworks, there are no other software bundles which provide means to support MPAD so far. The reason might be the fact that a standard for MPs does not yet exist, see Section \ref{sec:mission_profiles}.

Recently, Hirler et al. presented a theoretical model for reducing stressors in MPs to two single parameters i.e. \emph{effective stress level} and \emph{effective stress time}. Experiments showed that the reliability data obtained fit theoretical predictions within statistical variations \cite{hirler2017evaluation}.
%

\subsection{Knowledge- and Model-based Systems Engineering}
Knowledge-based systems support the explicit representation of knowledge in a domain and their exploitation through providing appropriate reasoning mechanisms \cite{tasso1998development}.

Sarder et al. explored the development of Systems Engineering ontologies in \cite{sarder2007developing}. An ontology for Systems Engineering has been proposed by van Ruijven in \cite{van2013ontology}. Based on the ISO 15926 standard, processes defined in the ISO 15288 standard are modeled. The approach does not directly support MPAD. Nevertheless, the ontology could be used for explicit description of the process presented in this paper using defined semantics.

In the Space Systems domain the application of ontologies in Model-Based Systems Engineering (MBSE) has been explored by Hennig et al. \cite{hennig2015languages,hennig2016ontology}. This approach uses ontologies to describe the system model and the conceptual data model and focuses on the integration of data models of various disciplines. It does not consider MPs and does not implement a CIA approach.
Ernadote also presented an approach to support MBSE combining standard meta models with dedicated ontologies to make understanding and creation of models easier \cite{ernadote2015ontology}. This approach hides the complexity of basic meta models and presents users ontology-based common vocabularies instead. These vocabularies are based on category-theory to support reading and creation of model data. The approach does not implement a CIA, does not provide any means for design parameter management and does not consider MPs.
%

\subsection{Ontologies in Requirements Engineering}
Ontologies in Requirements Engineering (RE) have been used to describe requirements specification documents or for the formal representation of requirements and application domain knowledge.They have also been used to check consistency and completeness \cite{castaneda2010use,siegemund2011towards}.

Dermeval et al. pointed out that most studies about the application of ontologies in RE focus on functional requirements \cite{dermeval2016applications}. Our approach can handle non-functional requirements, too. The literature review also revealed that only half of the studies followed W3C\footnote{https://www.w3.org/} recommendations on ontology-related languages.

Firesmith pointed out in \cite{firesmith2005your} that the definition of requirements meta data contributes to completeness. Although excluded from the scope of this paper, the use of ontologies in our approach allows specification of machine readable meta data using defined semantics thus enabling reasoning on such data. As Siegemund et al. stated in \cite{siegemund2011towards}, relationships among requirements are inadequately captured. Ontologies provide the means for formal representation of relations between concepts. As a result, they are suited to define and specify requirement relationships in Requirements Elicitation.
%
Our approach provides the means to lifting requirements descriptions to formal and semantically enriched representations and thereby has the potential to also support various RE activities. Moreover, the work of Siegemund et al. could be used to improve our approach with regard to the requirements representation.

There are various approaches to model requirements.
We found that most approaches use some kind of requirement model, although they do apparently not agree on a common type of model except for using ReqIF\footnote{http://www.omg.org/spec/ReqIF/} as a container format, for instance. Nevertheless, ReqIF can transport highly formalized as well as non-formalized requirements. This approach expects requirements to be represented in the most abstract form - natural language. These abstract descriptions are complicated to process and must be lifted to ontological representations first. We do not focus on the lifting itself but on establishing the foundation for semantic enrichment as for example done by Ferdinand et al. in \cite{ferdinand2004lifting}.
%

\subsection{Change Impact Analysis}
According to Li et al. CIA techniques can be divided into four perspectives: traditional dependency analysis, software repositories mining, coupling measurement and execution information collection \cite{li2013survey}. Although the presented CIA technique can be classified as dependency analysis it is not a traditional (code-based) approach as this paper will show.

Mostly and especially in software CIA, approaches focus on source code (65 \% according to Lehnert in \cite{lehnert2011review}) or other expressions such as natural language for instance.
Recently, Borg et al. presented an approach with a specific focus on CIA of software artifacts that are not source code in \cite{borg2017ciaRecomm}. This approach uses textual content of issue reports as basic information source.

In general, the level of abstraction in CIA can be different and thus not only programming languages but also Hardware Description Languages can possibly be considered even in language-dependent approaches. Lately, CIA for hardware designs has been explored for example by Ring et al. in \cite{ring2016change}. His work presented a framework for CIA that focuses on Change Management by detecting relationships between specifications using different levels of abstraction. The paper pointed out that CIA approaches for hardware designs need to take into account multiple levels of abstraction. Although this approach is applicable to hardware designs, it does not consider requirements or MPs.

To date, many CIA approaches mostly focus on functional changes to design artifacts. This might be the reason why the system impact remains underestimated and unnoticed in practice and analyses are incomplete \cite{rovegaard2008empirical}. The approach that is described in this paper can handle both functional and non-functional changes and has a strong focus on the analysis of overall system impact.

Please find more information on software CIA approaches in a detailed overview presented by Lehnert in \cite{lehnert2011review}.

\section{Approach}
\label{sec:approach}
This section describes the approach and provides an overview of the concept first. Then follows a explanation of how CIA can be realized applying a \emph{system model} that supports Trace Links and how MPs are integrated. The section concludes with a description of how design parameters are linked to requirements.

\subsection{Overview}
\begin{figure*}
  \centering
  \includegraphics[width=.95\linewidth]{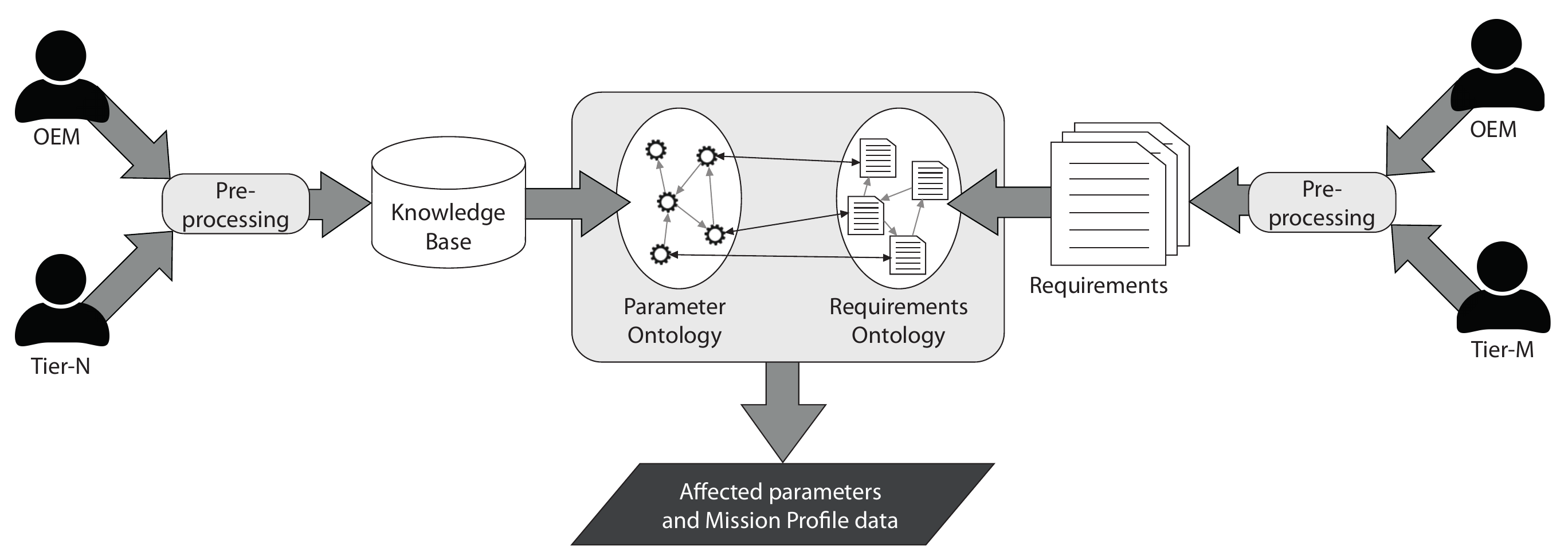}
  \captionsetup{width=.8\linewidth}
  \caption[bla]{Overview of the general approach: Expert knowledge about operative relationships between design parameters is lifted to a parameter ontology.  Requirements are lifted to a requirements ontology followed by the insertion of links between design parameters and requirements. A subsequent change affecting parameters can be determined and all relevant MP data can be passed to the analyst}
%
  \label{fig:approach_overview}
\end{figure*}
A core concept of the approach is the link between design parameters and requirements, see Fig. \ref{fig:requirements-passing-on}. In our approach, this is achieved through lifting the expert knowledge stored in a knowledge base and the requirements found in requirements specification documents to ontologies, see Fig. \ref{fig:approach_overview}. Once the ontological representations are ready, links are inserted semi-automatically, as described in Section \ref{sec:linking_requirements_to_design_parameters}. As soon as the system model as described in Section \ref{sec:system_model}, has been established and links to requirements have been inserted, standard graph analysis techniques can be used to determine parameters affected by a change. This basically means collecting all nodes on a path starting from a root node to all leaf nodes.

\subsection{Framework design}
\label{sec:semantic_mpad_platform}
This section describes the framework that was designed to provide the means to achieve MPAD tasks. The \emph{Semantic Mission Profile Aware Design Platform} served to accomplish the construction of the presented system but can and will also be used in other settings in the future.

Performing CIA involves connecting and reasoning on multiple heterogeneous data sources. To accomplish this task, using semantic technologies is beneficial as they encourage consolidation of different data sources and improved integration into the processes. The presented approach therefore uses a platform that has been created especially to support MPAD tasks for semantically enriched MPs. The conclusion of this section will give a brief description of the platform and outline the benefits from using it for the presented approach.

\subsubsection{Architecture and implementation}
To accomplish MPAD tasks, a comprehensive framework has been created - the \emph{Mission Profile Framework} (MPF). It allows developers to create and manage MP documents according to the MP format draft mentioned in Section \ref{sec:mission_profiles}.
The platform is implemented as a server application written in the Scala programming language \cite{odersky2004overview} and is meant to be used in conjunction with the MPF but can also be used as a stand-alone application. The platform realizes special MPAD tasks in the form of applications and extends resp. uses the tool and code base that the platform provides.

\subsubsection{Core functionality}
The extendable semantic MPAD platform is designed to process interlink and integrate MPs and requirements which just the MPF alone cannot do.
Input documents are therefore lifted to ontological representations by mapping engines. Resulting ontologies contain and allow for the definition of relationships between objects and thus favor semantic processing.
%

Apart from lifting documents to ontological representations, the platform provides means to integrate Change Management for requirements into workflows. This can be done using a tool that operates with ontological requirements specifications and is able to recognize changes made to them. Using the tool enables automatic reaction to changes and could also be extended to trigger certain actions based on the type of change, for instance.

\subsubsection{Interfaces}
The main intention in the design of a semantic MPAD platform was tight integration into existing design flows. Therefore, the platform provides a RESTful \cite{fielding2000architectural} API allowing users direct access to platform features through a web-based application with a HTML and JavaScript GUI. At the same time, the setup burden is reduced. Custom client applications which can fulfill additional special tasks when accessing the platform can use the same API. With regard to requirements, documents in ReqIF format containing requirements specifications are supported directly.
All resulting OWL ontologies are serialized to RDF/XML\footnote{https://www.w3.org/TR/rdf-primer/} and can be published to
user-defined triple stores via the SPARQL Protocol And RDF Query Language\footnote{https://www.w3.org/TR/sparql11-overview/} or be accessed directly through the platform server API.

\subsection{System model}
\label{sec:system_model}
The system model serves the purpose of specifying various relevant relationships, see Fig. \ref{fig:system_model} for an example. This model is a means to specify knowledge about the system for model integration, insert Trace Links and support the CIA, as explained in Section \ref{sec:change_impact_analysis}. It is used to specify the relationships
\begin{itemize}
\item between \emph{components} and \emph{functions}
\item between \emph{components} and \emph{design parameters}
\item between \emph{components} and \emph{Mission Profiles}
\end{itemize}
and for the implementation of Requirements Traceability
\begin{itemize}
\item between \emph{systems} and \emph{requirements}
\item between \emph{components} and \emph{requirements}.
\end{itemize} 
The system model is expressed as an OWL ontology and is meant to ease model integration by referring to a common uniform model. The primary purposes are:

\subsubsection{Supporting Change Impact Analysis}
\label{sec:change_impact_analysis}
The CIA is primarily enabled through specification of relationships between \emph{requirements} and \emph{design parameters} using the system model. These relationships represent Trace Links. We propose a semi-automatic procedure in order to collect these links: candidates for links are suggested to the user who can either accept or decline them. See Section \ref{sec:linking_requirements_to_design_parameters} for a detailed description of this methodology.

\subsubsection{Enabling Requirements Traceability}
The system model can be used to enable Requirements Traceability through insertion of Trace Links as mentioned in the previous section.

\subsubsection{Integrating Mission Profiles}
\begin{figure}
  \centering
  \includegraphics[width=\linewidth]{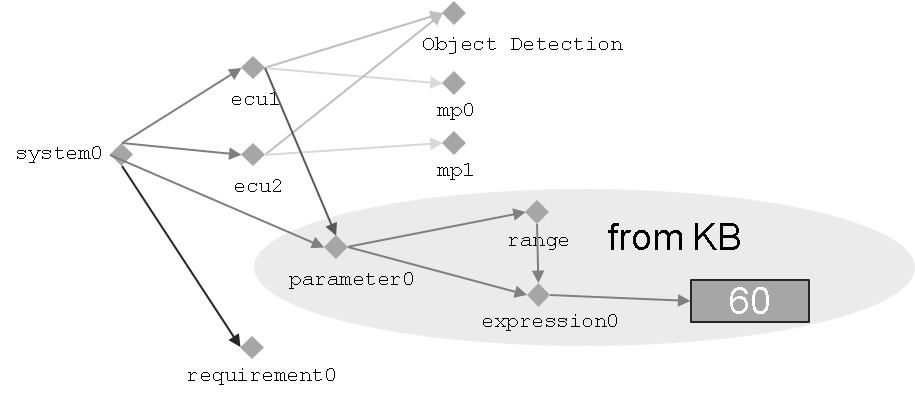}
  \caption[bla]{System model example instance modeling a single instance of a system composed of two Electronic Control Units (ECUs), which realize the \emph{Object Detection} function. The two ECUs are linked to corresponding MPs. One of the ECUs is also linked to a \emph{design parameter
  } which specifies that the \emph{range} of the component is "60". While the parameter and it's expression stem from the knowledge base, the link between the ECU component and the parameter is user-specified}
  \label{fig:system_model}
\end{figure}

MP data is integrated by inserting links between components and MP document representations in the system model, see also Fig. \ref{fig:system_model}. Ontological MP document representations are used and can be queried for specific data portions.

\begin{algorithm}
\SetAlgorithmName{Alg.}
\SetAlgoLined{}
\KwData{requirement description $r_i \in \mathcal{R}$, parameter description $p_j \in \mathcal{P}$, $i,j \in \mathbb{N}_0$, $0 \leq i \leq |\mathcal{R}|$, $0 \leq j \leq |\mathcal{P}|$}
 \KwResult{set of link suggestions $\mathcal{L}$}
\ForEach{$r_i \in \mathcal{R}$}{
\ForEach{$p_j \in \mathcal{P}$}{
 tokenize $r_i$ and $p_j$ \;
 part-of-speech tagging of all tokens \;
 recognize lemmas of all tokens \;
 \ForEach{token $t$ of tokens of $r_i$}{
 \ForEach{token $s$ of tokens of $p_i$}{
 \If{$t$ and $s$ are nouns}{
 get lemma $l$ of $t$ and $k$ of $s$ \;
 get the set $\mathcal{S}$ of symmetric relationships between $l$ and $k$ \;
 \If{$\mathcal{S}$ is not empty}{
 add an entry to the set of link suggestions $\mathcal{L}$ with a link between $r_i$ and $p_i$ \;
 }
 }
 }
 }
}
}
\captionsetup{width=.8\linewidth}
\caption{Link suggestion}
\label{alg:link_suggestion}
\end{algorithm}

\subsection{Linking requirements to design parameters}
\label{sec:linking_requirements_to_design_parameters}
Design parameters need to be linked to corresponding requirements in oder to use gathered expert knowledge about operative relationships. For this purpose, the proposed system  contains a component for the specification of such links and, what is even more important, a component which is able to suggest links to the user providing a semi-automatic way to specify these links.

\subsubsection{Specification of links}
All relevant knowledge (system model, requirements, design parameters) is expressed via ontologies, following W3C recommendations. With regard to the underlying RDF data model, each design parameter or requirement can be seen as a \emph{resource}.
A link between these is then represented through a triple connecting both by a predicate. We identified basically two options for the specification of these links:

(1) indirect specification by generating individuals representing the link:
\begin{lstlisting}[frame=none,label=lst:explicit_link,xleftmargin=3.4pt,xrightmargin=3.4pt]
ex:link1 rdf:type ex:Link
ex:link1 ex:hasReq req:requirement1
ex:link1 ex:hasDp dp:designParameter1
\end{lstlisting}
(2) direct specification by adding a relationship between the requirement and the design parameter instances:
\begin{lstlisting}[frame=none,label=lst:implicit_link,xleftmargin=3.4pt,xrightmargin=3.4pt]
req:requirement1 dpr:relatedTo dp:designParamter1
\end{lstlisting}
The difference between these representations is above all that it is easier to add information to a link in the first form. This is the reason why we decided to represent links using this form.
%

\subsubsection{Suggestion of links}
To support users in the process of specifying links between requirements and design parameters, the proposed system contains a component which is able to suggest link candidates to the user, see Fig. \ref{fig:system_component_diagram}. The user can either accept or decline a link suggestion. Requirements as well as design parameters are expected to be descriptions in natural language. We use Stanford CoreNLP \cite{manning2014stanford} for natural language processing tasks.

The main idea behind our examplary link suggestion is the use of WordNet \cite{miller1995wordnet} to identify related synonyms in nouns of natural language requirement respectively design parameter descriptions.
This task can be achieved for example with Alg. \ref{alg:link_suggestion}. Although performance could be optimized and it could be extended to compare also the string distances of nouns in lines 8-14 to handle typos in descriptions, for instance, it can be used to find candidates for links in certain situations.

\section{Results}
\label{sec:use_case}
In this section we first describe a framework which was used to capture expert knowledge. Then, we employ our approach to analyze the impact of changes in two use cases.
The first use case is based on the power consumption of logic gates and registers model in the TAMTAMS \cite{vacca2012tamtams} tool. In this use case we focused on the impacted parameters. The second use case is a real-time application in which the real-time requirements are transformed into technology-level requirements. 
Focus of the second use case was deriving requirements of parameters on a lower level.

\subsection{Capturing expert knowledge}
\begin{figure*}
  \centering
  \includegraphics[width=.8\linewidth]{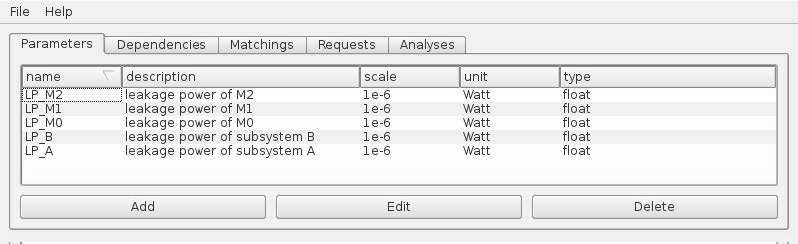}
  \captionsetup{width=.8\linewidth}
  \caption[bla]{CTEF client application GUI showing the parameter specification tab which offers the possibility to add, edit and remove parameter definitions. Other tabs include dependency, matching and request management functions as well as utilities for conducting analyses}
  \label{fig:ctef-client}
\end{figure*}

The \emph{Collaborative Technology Evaluation Framework} (CTEF) provides the means to capture expert knowledge about operative relationships and was designed specifically for this task.
It represents the knowledge base component in the general approach, see Fig. \ref{fig:approach_overview}.
The centralized system currently contains a server and a client which are loosely coupled and are therefore exchangeable. While the server could for example be hosted by a trusted neutral party, the clients are meant to be used cross-domain by all participants in a supply chain. Fig. \ref{fig:ctef-client} shows a screenshot of the CTEF client application.

\subsubsection{Dependencies and matchings of parameters}
Gathered knowledge is stored on the server-side and is represented as a \emph{Dependency Matching Graph} (DMG). Apart from the specification of design parameters which can be set to be either \emph{private} or \emph{public}, CTEF allows for definition of dependencies between design parameters. The definition of relationships between parameters is a fundamental concept of the framework. As this a collaborative framework, these relationships can be either \emph{dependencies} or \emph{matchings}. Matchings are relationships from parameters to parameters of providers. The idea behind is that a component or system might have some top level parameters e.g. the range of a radar system which are themselves depending on other parameters, which need to be supplied by a different provider. A radar system might for example have an amplifier component manufactured by a another provider in the supply chain. As the range of a radar system also depends on the gain of its amplifier, the manufacturer of the amplifier component will also be the \emph{provider} of a corresponding design parameter. A pair of corresponding parameters is a \emph{match} and together with the dependencies forms the DMG, see Fig. \ref{fig:tamtams_dmg} for a visualization of a DMG.

\subsubsection{Transferring knowledge}
\label{sec:ctef_transferring_knowledge}
To actually use the knowledge gathered with CTEF the server component has on the one hand functionality which allows analyzing various aspects of DMG relations and is on the other hand able to export captured information in RDF/XML format. Using the RDF data model constitutes the interface to knowledge bases and systems such as the platform described in Section \ref{sec:semantic_mpad_platform} and is the basis for semantic enrichment and reasoning via the creation of OWL ontologies. In addition, this allows the performance of semantic queries for example with SPARQL against captured knowledge. Introducing this Web-compliance feature allows for interlinking knowledge across deployed CTEF systems on Web-scale to discover and explore most complex operative relationships.

\subsection{Use case: Power consumption}
\begin{figure*}
  \centering
  \includegraphics[width=.8\linewidth]{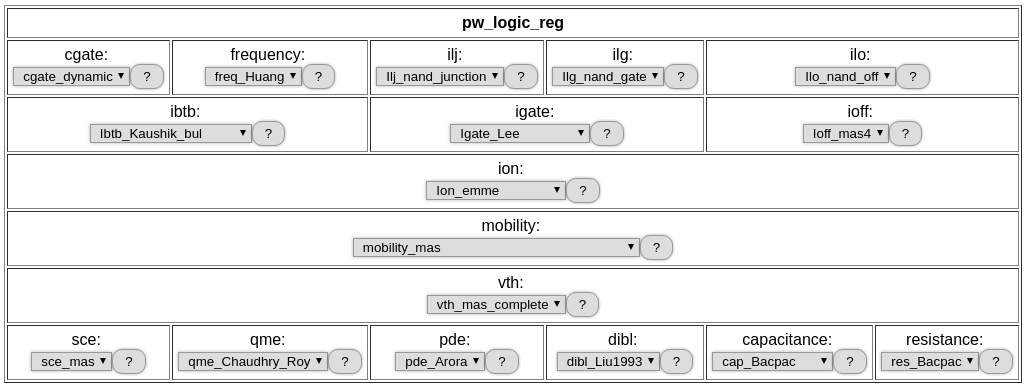}
  \captionsetup{width=.8\linewidth}
  \caption[bla]{Web GUI of TAMTAMS for the analysis of the power consumption of logic gates and registers}
  \label{fig:tamtams}
\end{figure*}

\begin{figure}
  \centering
  \includegraphics[width=\linewidth]{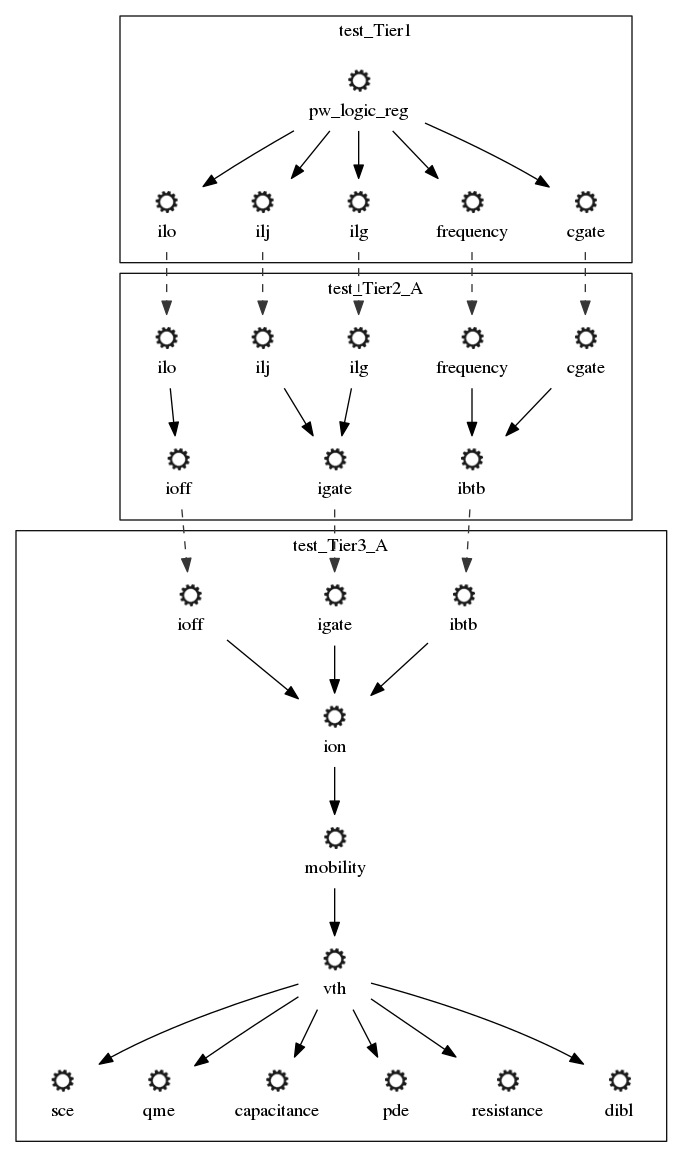}
  \caption[bla]{Translation of the implicit graph structure of the power consumption of logic gates and registers analysis of TAMTAMS to a DMG in CTEF}
  \label{fig:tamtams_dmg}
\end{figure}

\begin{figure}
  \centering
  \includegraphics[width=\linewidth]{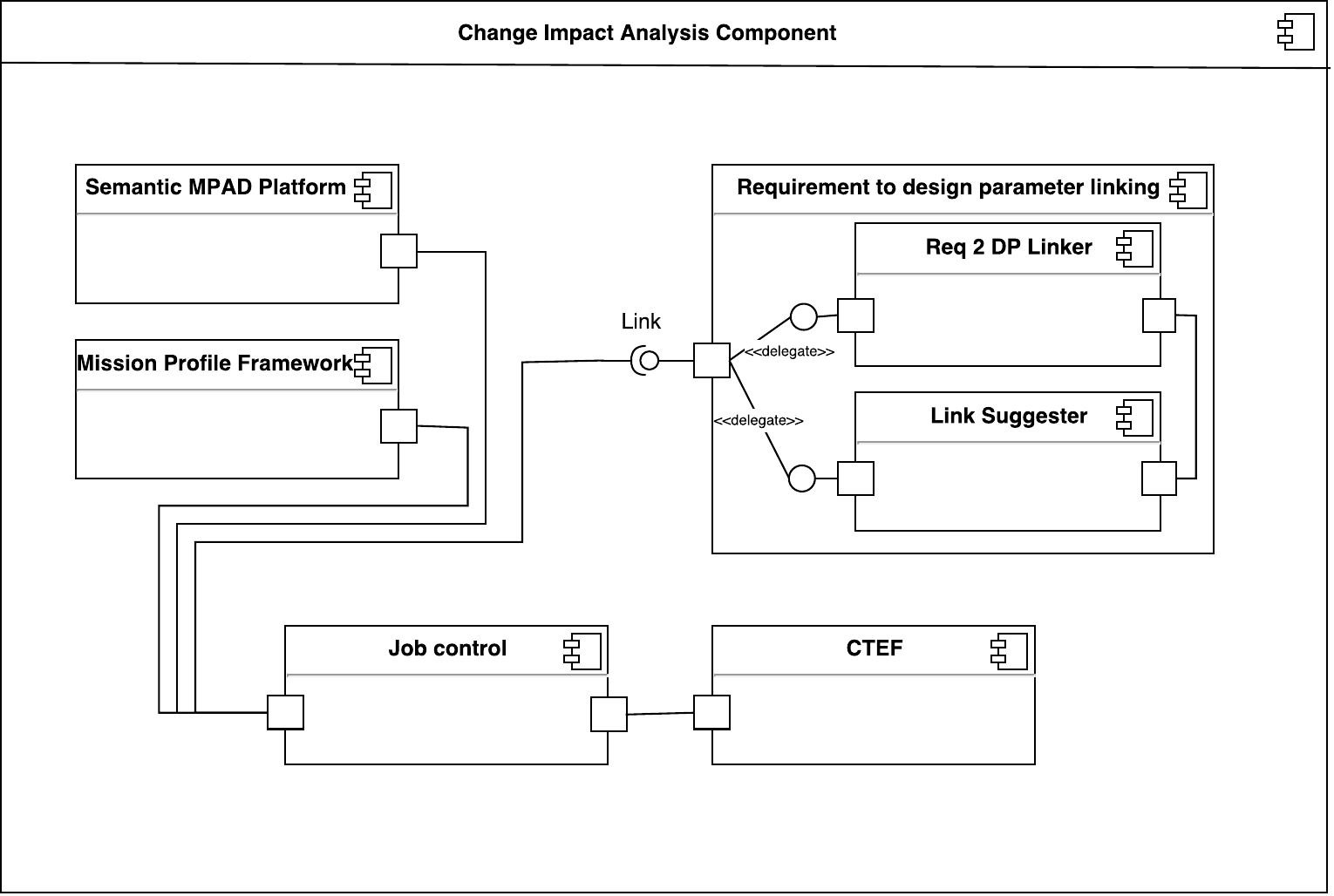}
  \caption[bla]{UML Component diagram of the system realizing the presented approach. All components are connected and controlled by a central control component. Please notice the interface to the \emph{requirement to design parameters linking} component, which is introduced to allow for further experimenting with different approaches in linking}
  \label{fig:system_component_diagram}
\end{figure}

The TAMTAMS module used, is the system level module for analyzing power consumption of logic gates and registers. This module consists of several low level technology parameters, as well as high level parameters such as the frequency or the overall power consumption, see Fig. \ref{fig:tamtams}.
The dependencies of the parameters form an implicit graph which is explicitly specified using CTEF, see Fig. \ref{fig:tamtams_dmg}.
Please refer to Fig. \ref{fig:system_component_diagram}, for an overview of the system components and how they are connected.
The grouping of parameters into tiers is just an example and can of course be different in practice. However, this does not affect the approach's outcome. The relationships between design parameters are important and their formation of a hierarchy with design parameters residing on various levels. The set of design parameters and the relationships between them is the knowledge gathered from domain experts via CTEF. In this example, the knowledge about the operative relationships between design parameters is taken from TAMTAMS.

Apart from expert knowledge about operative relationships between design parameters gathered through CTEF, knowledge about requirements is lifted to corresponding ontologies, too. Once the ontologies are ready, the system starts to ingest knowledge about the relationships between requirements and design parameters from user input. The developers establish the system model connecting  knowledge and forming the main knowledge corpus. Links between components and MPs are inserted.

Consider the case that the requirement that specifies the system frequency is changed during the development. This is detected as a change to a ReqIF document containing requirements of the system which is being developed. The system resolves the dependencies of the changed requirement to find out which other requirements are affected. As a consequence, the system can detect which design parameters are affected by the change. This is achieved by using the knowledge about operative relationships between design parameters exported as ontology from CTEF and being connected to the system model.
Finally, the result is presented to the analyst - affected design parameters: \emph{frequency, ibtb, ion, mobility, vth, sce, qme, pde, dibl, capacitance} and \emph{resistance}.

\subsection{Use case: Real-time application}
For further evaluation of the approach we applied it to a traffic sign recognition application which is widely used in automated driving systems. It requires a real-time processing of frames and it runs on an embedded platform. Since there are 24 frames per second there is a real-time requirement which requires that each frame is processed in \mbox{42 ms}. This requirement can be translated to a requirement of write latency of the memory system.

We trained a model to map the processing time of a single frame to the write latency of the main memory.
%
This was realized by running a vast variety of different benchmarks\footnote{http://www.mrtc.mdh.se/projects/wcet/benchmarks.html} measuring the performance counters such as the number of cache misses, predictable branches and data read accesses.
The model is build up using machine learning to map the performance counters and the write latency to the processing time of a single frame. The results were extracted using the ARM based Lauterbach Trace32 on the Xilinx Zedboard with a Zynq-7000 SoC board equipped with a dual-core Cortex-A9.
Based on the processing requirement of 42 ms per frame at 600 Mhz the write latency should not exceed 31 cycles.

Using our proposed system, we can link this requirement to corresponding design parameters. The proposed system also allows specification of dependencies between requirements and taking them into consideration.

\section{Conclusion}
\label{sec:conclusion}
The paper focused on the identification and specification of links between requirements and design parameters to connect different knowledge sources for the purpose of establishing a common system model containing system knowledge and Trace Links to support CIA. There are possibilities to improve the approach by extending it with methodologies found in the literature, e.g. repository mining for the consideration of the history of changes.
The presented approach expects all input to be in the most abstract representation - natural language. If models were used instead, this might ease linking of artifacts as models could be queried for specific attributes used for the decision whether two artifacts relate to each other. This also leads to another opportunity for an improvement: the application of Ontology Mapping or Entity Matching approaches to link requirements to design parameters which is crucial to the practical applicability of our approach.
%

Further research could be carried out with regard to how reasoning could be used to check for consistency or even to identify links between requirements and design parameters. Important is also the lifting of natural language requirements to formal specifications which is a prerequisite to our approach. It might also be reasonable to explore ways of integration of the presented approach into Requirements Engineering activities such as Requirements Management. Here, the approach could be used to ensure completeness by pointing out missing requirements of design parameters, which was identified as the most difficult task within Requirements Analysis \cite{davis1990software}.





\section{Acknowledgement}
This paper is partially supported by the BMBF projects autoSWIFT (grant number 16ES0358), SAFE4I (grant number 01\textbar{}S17032C) and the State of Baden-Württemberg, Germany, Ministry of Science, Research and Arts within the cooperative graduate program EAES of the University of Tübingen.

%
%

\balance

\bibliographystyle{IEEEtran}
\bibliography{main}

\end{document}